\documentstyle[epsf]{mn}

\begin{document}
\def\lsun{{\rm L_{\odot}}}
\def\msun{{\rm M_{\odot}}}
\def\rsun{{\rm R_{\odot}}}
\def\be{\begin{equation}}
\def\ee{\end{equation}}

\title{Cygnus X-2, super-Eddington mass transfer, and pulsar binaries}
\author[A. R. King \& H. Ritter]{A. R. King$^1$
\& H. Ritter$^{2}$\\ 
$^1$ Astronomy Group, University of Leicester,
Leicester, LE1~7RH\\ $^2$ Max--Planck--Institut f\"ur Astrophysik, 
Karl--Schwarzschild--Str. 1, D--85740 Garching, Germany}

%-----------less/greater than approx eq to--------------
\newcommand{\lta}{\la}
\newcommand{\gta}{\ga}
%-------------------------------------------------------

\maketitle
\begin{abstract}
We consider the unusual evolutionary state of the secondary star in
Cygnus X-2. Spectroscopic data give a low mass
($M_2 \simeq 0.5 - 0.7\msun$)
and yet a large radius ($R_2 \simeq 7\rsun$) and high luminosity ($L_2
\simeq 150\lsun$). We show that this star closely resembles a remnant of
early massive Case B evolution, during which the neutron star ejected most
of the $\sim 3\msun$ transferred from the donor (initial mass 
$M_{\rm 2i}\sim 3.6\msun$) on its thermal time-scale $\sim 10^6$~yr.
As the system is far too
wide to result from common-envelope evolution,
this strongly supports the idea that a neutron star efficiently
ejects the excess inflow during super--Eddington mass transfer.
Cygnus X-2 is unusual in having had an
initial mass ratio $q_{\rm i} = M_{\rm 2i}/M_1$ in a narrow critical range near
$q_{\rm i}\simeq 2.6$. Smaller $q_{\rm i}$ lead to long-period
systems with the former donor near the Hayashi line, and larger $q_{\rm i}$
to pulsar binaries with shorter periods and
relatively massive white dwarf companions. The latter
naturally explain the surprisingly large
companion masses in several millisecond pulsar binaries.
Systems like Cygnus X-2 may thus be an important channel for forming
pulsar binaries.

\end{abstract}

\section{INTRODUCTION}
\label{sec:intro}
Cygnus X-2 is a persistent X-ray binary with a long orbital period
($P = 9.84$~d, Cowley, Crampton \& Hutchings 1979). The observation of
unambiguous Type I X-ray bursts (Smale, 1998) shows that the accreting
component is a neutron star rather than a black hole. The precise 
spectroscopic information found by Casares, Charles \& Kuulkers
(1998), and the parameters which can be derived from it, is summarized
in Table 1. The mass ratio $q = M_2/M_1 \simeq 0.34$ implies that mass
transfer widens the system, and is therefore probably 
driven by expansion of the secondary star. Normally in long-period
low-mass X-ray binaries (LMXBs) this occurs because of the nuclear
evolution of a subgiant secondary along the Hayashi line, with typical
effective temperatures $T_{\rm eff,2} \simeq 3000 - 4000$~K.
However Casares et al.'s observations show that this cannot be the
case for Cygnus X-2. The secondary is in the Hertzsprung gap
(spectral type A9 III): use of Roche geometry and the
Stefan--Boltzmann law gives 
$L_2 \simeq 150\lsun$ with $T_{\rm eff,2} \simeq 7330$~K (see Table 1). 
Moreover the mass ratio $q \simeq 0.34$, and the assumption that
the primary is a neutron star and thus obeys $M_1 \la 2\msun$, implies that 
the secondary has a low mass ($M_2 = qM_1 \la 0.68\msun$). In
contrast, an isolated A9 III star would have a
mass of about $4\msun$. More recently Orosz \& Kuulkers (1998) have
modelled the ellipsoidal variations of the secondary and thereby
derived a model-dependent inclination of $i = 62.5^{\circ} \pm
4^{\circ}$ which translates into component masses $(M_1 = 1.78
\pm 0.23) \msun$ and $(M_2 = 0.60 \pm 0.13) \msun$. 

In this paper we consider explanations for the unusual nature
of the secondary in Cygnus X-2. We find only one viable possibility,
namely that this star is currently close to the end of early
massive Case B mass transfer, and thus that the neutron star has
somehow managed to reject most of the mass ($\sim 3\msun$) transferred
to it in the past. In support of this idea, we show that this type of
evolution naturally explains the surprisingly large companion
masses in several millisecond pulsar binaries. 
\begin{table}

\caption{Observed and derived properties of Cygnus X-2.} 
%The
%derivations assume that the absence of eclipses restricts the orbital
%inclination to $i \la 73^{\circ}$} 
%and that the primary is a neutron
%star with $M_1 \la 2\msun$.} 

\begin{tabular}{ll}
spectroscopic period $P$ 	        &	$9^{\rm d}.844(\pm 3)$ \\
radial-velocity amplitude $K_2$      & $88.0\pm 1.4$ km s$^{-1}$ \\
rotational velocity $v_2\sin i$       & $34.2 \pm 2.5$ km s$^{-1}$ \\
spectral type Sp(2)                   & A$9 \pm 2$ III \\
%\end{tabular}
%
%\end{table}
%
%\begin{table}
%
%\caption{Derived properties of Cygnus X-2}
%
%\begin{tabular}{ll}
mass ratio $q = M_2/M_1$                         & $0.34\pm 0.04$ \\
effective temperature $T_{\rm eff,2}$            & $(7330\pm 320)$~K \\
primary mass $M_1$                               & $\ga (1.43\pm
0.10)\msun$ \\
secondary mass $M_2$                             & $\ga (0.49\pm
0.07)\msun$ \\
                                                 & $\la (0.68\pm
0.08)\msun$ \\
secondary radius $R_2$                           &
$8.93\rsun \left( \frac{M_2}{\msun} \right)^{1/3}$ \\
secondary luminosity $L_2$                       & $207.7\lsun
\left( \frac {M_2}{\msun} \right)^{2/3}
\left( \frac {T_{\rm eff,2}}{7330~{\rm K}}\right)^4$ \\    
\end{tabular}

\end{table}

\section{MODELS FOR CYGNUS X-2}

In this Section we consider four possible explanations for the
unusual nature of the secondary in Cygnus X-2. We shall find that
three of them are untenable, and thus concentrate on the fourth 
possibility.

\subsection{A normal star at the onset of Case B mass transfer?}
\label{a}

The simplest explanation is that the position of the secondary in the
HR diagram is just that of a normal star crossing the Hertzsprung
gap. Since such a star no longer burns hydrogen in the core, this is
a massive Case B mass transfer as defined by Kippenhahn \& Weigert
(1967), hereafter KW. 
Provided that the initial mass
ratio $q_{\rm i} \lta 1$  the binary always expands on mass transfer,
which occurs on a thermal time-scale.
Kolb (1998) investigated this type of evolution
systematically and found that the secondary's position on the HR
diagram is always close to that of a single star of the same
instantaneous mass. For Cygnus X-2 the A9 III spectral type
would require a current secondary mass $M_2 \simeq 4\msun$, and thus 
a primary mass $M_1 = M_2/q \simeq
12\msun$. This is far above the maximum mass for a neutron star and
would require the primary to be a black hole, in complete
contradiction to the observation of Type I X-ray bursts from Cygnus X-2
(Smale, 1998). 

We conclude that the secondary of Cygnus X-2 cannot be a normal
star. Accordingly we 
must consider explanations in which it is far bigger and more
luminous than expected from its estimated mass $\sim (0.49 - 0.68)\msun$.

\subsection{A stripped subgiant?}

The type of Case B evolution described in subsection (\ref{a}) above is known
as `early', in that mass transfer starts when 
the donor's envelope is still largely
radiative rather than convective (as it would become as the star
approached the Hayashi line), and `massive', meaning that the 
helium core has a large enough mass that upon 
core contraction it does not become degenerate but instead 
ignites central helium burning. The corresponding minimum core 
mass is about $0.35 \msun$, corresponding to a total ZAMS mass of 
$2 - 2.5\msun$ (depending on the assumed degree of convective
overshooting during central hydrogen burning). For lower initial
masses we have `low-mass' Case B (Kippenhahn, Kohl \& Weigert 1967). Here
the donor's helium core becomes degenerate and 
the envelope is convective. After a possible early rapid mass transfer
phase in which the binary mass ratio $q = M_2/M_1$ is reduced to $\la 1$,
(Bhattacharya \& van den Heuvel, 1991; Kalogera \& Webbink, 1996)
the donor reaches the Hayashi line and
mass transfer is driven by its nuclear expansion
under hydrogen shell burning. The star remains on the Hayashi
line, increasing its radius and the binary period as its core mass
grows. Webbink, Rappaport \& Savonije (1983) describe this type of
`stripped giant/subgiant'
evolution in detail, and indeed fit Cygnus X-2 in this
way. However there is a clear discrepancy between the 
observed effective temperature and that required for a Hayashi-line
donor, which
should be about 4100 K in this case. Webbink et al. (1983) appeal to
X-ray heating by the primary to raise the temperature to the observed
7330~K, but remark that since the heating only operates on the side
of the secondary facing the primary, one would expect a much larger
orbital modulation of the optical flux than actually observed ($\Delta
V_{\rm obs} \simeq 0.3$~mag) unless the orbital inclination $i$ is
low. Simple estimates (see the Appendix) 
show that such a small modulation would require
$i < 13.4^{\circ}$.
However the mass function and mass ratio for Cygnus X-2 can be
combined to show that $M_1\sin^3i = (1.25\pm 0.09)\msun$, 
so such small inclinations would imply $M_1 > 100\msun$, again clearly
incompatible with the very strong 
observational evidence for a neutron star primary. A still stronger
argument can be constructed on the basis that the spectral type of the
secondary is not observed to vary during the orbital cycle.

\subsection{A helium white dwarf undergoing a hydrogen shell flash?}

It is known that newly-born helium white dwarfs 
can undergo one or more hydrogen
shell flashes during their evolution from the giant branch
to the white dwarf cooling sequence. During these flashes the star has
a much larger photosphere. Calculations by Driebe et al. (1998) show
that 
only low-mass He white dwarfs in the interval $0.21\msun \lta M_{\rm WD} \lta
0.30\msun$ can undergo such a flash which in turn can put the star in 
the same position on the HR diagram as the secondary of Cygnus X-2
(e.g. the first shell flash of the sequence with $M_{\rm WD} =
0.259\msun$). However, the required
low secondary mass has a price: the observed mass ratio 
implies $M_1 = qM_2 \simeq (0.76\pm 0.09)\msun$, much smaller than
required by the mass function ($M_1\sin^3i = (1.25\pm 0.09)\msun$). To
makes matters worse, the evolutionary track crosses the relevant
region of the HR diagram in an extremely short time: the star's radius
expands on a time-scale $\tau = {\rm d}t/{\rm d}\ln R_2 \simeq
33$~yr. Not only does this give the present system an implausibly
short lifetime, the radius expansion would drive mass transfer at a
rate $\sim \Delta M_{\rm H}/\tau \sim$ few $\times 10^{-4}\msun$~yr$^{-1}$,
again totally inconsistent with observations. We conclude that the
secondary of Cygnus X-2 cannot be a low-mass He white dwarf
undergoing a hydrogen shell flash.

\subsection{A star near the end of early massive Case B mass transfer?}

We saw in subsection (\ref{a}) above that a secondary near the {\it
onset} of early massive Case B mass transfer
(i.e. with $q \lta 1$ throughout) is ruled out for
Cygnus X-2, as the required stellar masses conflict with observation. 
However, a more promising assignment is a secondary near the {\it end}
of an early massive Case B evolution which began with $q_{\rm i} \ga 1$.

KW have investigated this process in detail.
In contrast to the case $q_{\rm i} \lta 1$ discussed by Kolb (1998), and 
considered in (\ref{a}) above, the ratio $q_{\rm i} \gta 1$ 
means that the binary and Roche lobe initially shrink on mass transfer.
Adiabatic stability is nevertheless ensured because the
secondary's deep radiative envelope (`early' Case B) contracts on
rapid mass loss. Mass transfer
is therefore driven by the thermal-time-scale expansion of the
envelope, but is more rapid than for $q_{\rm i} \lta 1$ because of the
orbital shrinkage. Once $M_2$ is reduced to the point that
$q \lta 1$, the Roche lobe begins to expand. This slows
the mass transfer, and shuts it off entirely when the lobe reaches
the thermal-equilibrium radius of the secondary, since the latter then has
no tendency to expand further (except possibly on a much longer
nuclear time-scale).
Calculations by KW and Giannone, Kohl \& Weigert (1968), hereafter GKW, 
show that in some cases the orbit can shrink so much that the process
ends in the complete exhaustion of the donor's
hydrogen envelope, ultimately leaving the core of the secondary in a
detached binary. Alternatively, the rapid Case B mass transfer may
end with the donor on the Hayashi line, still retaining a large
fraction of its original hydrogen envelope. 
However, there will be no long-lasting phase of mass transfer with
the donor on the Hayashi line because the star starts shrinking with 
ignition of central helium burning (KW). 
Neither of these two cases describes Cygnus X-2.
However, there is an intermediate possibility: the initial mass ratio 
$q_{\rm i}$ may be such that the
donor retains a small but non-negligible hydrogen envelope as mass transfer
slows. The current effective temperature of 7330~K shows that the
companion's envelope is mainly radiative, with only a very thin
surface convection zone. (A paper in preparation by Kolb et al. shows
this in detail.)
In the example computed by KW the donor, at the end of mass
transfer, is not on the Hayashi line, but 
at almost the same point in the HR diagram as it occupied
immediately before mass transfer began (this is confirmed by the
detailed numerical calculations of Kolb et al.).
Just before the process ends 
we then have an expanding low-mass donor,
driving a modest mass transfer rate in a long-period
expanding binary, but at the
HR diagram position of a much more massive normal star. 
As we shall show below, for an initial donor mass of about $3.6\msun$ 
the end point of such an evolution can be made to match closely that
observed for the secondary of Cygnus X-2. (Note that we have not
performed detailed numerical calculations for this paper, but rather
used the results of KW and GKW. The forthcoming paper by Kolb et
al. reports detailed calculations.)

Clearly this idea offers a promising 
explanation of the secondary in Cygnus X-2. However there is an
obvious difficulty in accepting it immediately. KW's calculations 
assumed that the total binary mass and angular
momentum were conserved, and in particular that the primary retained
all the mass transferred to it. But the primary of Cygnus X-2 is known to
be a neutron star, with a mass presumably $\la 2\msun$,
so we must require instead that it accretes
relatively little during mass transfer. This agrees with the idea that
a neutron star cannot accrete at rates greatly in excess of its
Eddington limit ($\sim 10^{-8}\msun$~yr$^{-1}$), and the fact that
almost all of the mass is transferred at much
higher rates ($\ga 10^{-6}\msun$~yr$^{-1}$). 
We thus follow earlier authors 
(Bhattacharya \& van den Heuvel, 1991; Kalogera \& Webbink, 1996)
in postulating that the
neutron star is extremely efficient in ejecting most of the
super--Eddington mass transfer, 
rather than allowing the excess mass to build up
into a common envelope.
Clearly common-envelope evolution could not
produce Cygnus X-2: the current binary period of 9.8~d shows that far
too little orbital energy could have been 
released to remove the envelope of any plausible progenitor for the
secondary (see e.g. Section 4 below).
By contrast, it is at least energetically possible to expel most of
a super--Eddington mass transfer 
rate, provided that this is done at large enough
distance $R_{\rm ej}$ from the neutron star. If the matter is given
just the escape velocity at $R_{\rm ej}$ the ratio of ejection to
accretion rate is 
\be
{\dot M_{\rm ej}\over \dot M_{\rm acc}} ={R_{\rm ej} \over R_*},
\ee 
where $R_*$ is the radius of the neutron star. Ejecting all but about
1\% of the transferred matter thus requires 
\be
R_{\rm ej} \ga  100R_* \sim 10^8\ {\rm cm}
\label{ej}
\ee
which is far smaller than the size of the accretion disc around the
neutron star for example. (This point is discussed further by King and
Begelman, 1999.)

In the next section we will consider an early massive Case B 
evolution for Cygnus X-2. We will find that the hypothesis
of efficient mass ejection by the neutron star allows excellent
agreement with the current state of the system, as well as plausible
explanations for the observed states of several detached pulsar binaries.

\section{EARLY MASSIVE CASE B EVOLUTION FOR NEUTRON-STAR BINARIES}

\protect\label{mcb}
The main features of early massive Case B evolution can be
understood by considering the relative expansion or contraction of the
donor's Roche lobe $R_{\rm L}$ and the 
thermal equilibrium radius $R_{\rm 2,e}$ which the
donor attains at the end of mass transfer (see e.g. GKW).
As discussed above, we
assume that the neutron star ejects any super--Eddington mass inflow.
Since the mass transfer rate exceeds the Eddington limit by
factors $\ga 100$ (see above), almost all of the transferred mass
must be ejected, and to an excellent approximation 
we can assume that the neutron star mass $M_1$ remains fixed during
the mass transfer (the equation for $R_{\rm L}$ can actually be integrated
exactly even without this assumption, but at the cost of some
algebraic complexity). We assume further that the ejected mass carries
the specific angular momentum of the neutron star's orbit. This is
very reasonable, since the ejection region is much smaller than the
size of the disc (see eq. \ref{ej}). Then we can
use the result quoted by Kalogera \& Webbink (1996) to write
\be
{R_{\rm L}\over R_{\rm L,i}} = \biggl({M_{\rm 2i}\over M_2}\biggr)^{5/3}
                     \biggl({M_{\rm i}\over M}\biggr)^{4/3}
                     e^{2(M_2 - M_{\rm 2i})/M_1},
\protect\label{rl}
\ee
where $M_2$ and $M = M_1+M_2$ are the donor and
total binary mass at any instant, and $M_{\rm 2i}, M_{\rm i}$ their values at
the onset of mass transfer. In writing (\ref{rl}) we have used the
simple approximation 
$R_{\rm L}/a \propto (M_2/M)^{1/3}$, where $a$ is the binary
separation. Using this and 
Kepler's law we get the change of binary period $P$ as
\be
{P\over P_i} = \biggl({R_{\rm L}\over R_{\rm L,i}}\biggr)^{3/2}
               \biggl({M_{\rm 2i}\over M_2}\biggr)^{1/2}
\ee
so that  
\be           
{P\over P_i}
= \biggl({M_{\rm 2i}\over M_2}\biggr)^3
                     \biggl({M_{\rm i}\over M}\biggr)^2
                     e^{3(M_2 - M_{\rm 2i})/M_1}.
\protect\label{orb}
\ee
Conventional massive Case B
evolution always begins with a mass ratio $q_{\rm i} = M_{\rm 2i}/M_1$
sufficiently large that $R_{\rm L}$ initially shrinks. From (\ref{rl}) it is
easy to show that this requires $q_{\rm i} > 1.2$.
If during the evolution $M_2$ decreases
enough that $q < 1.2$, $R_{\rm L}$ begins to expand again. The curves of
$\log R_{\rm L}$ thus have the generic U-shaped forms shown in
Figs. 1 -- 3.

\begin{figure}
  \centerline{\epsfxsize=1.0\hsize\epsffile{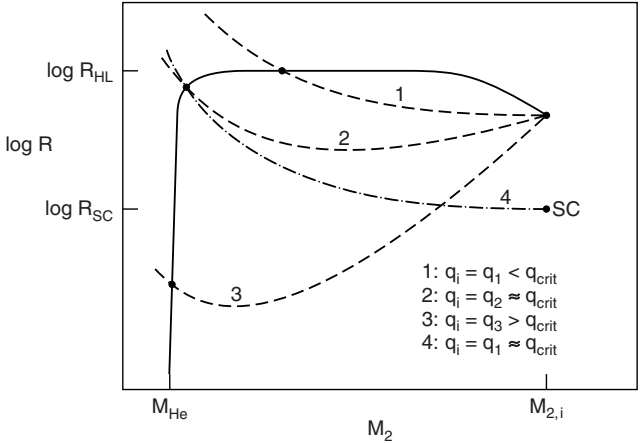}}
\caption{Schematic mass radius diagram of early massive Case B evolution 
showing different outcomes of mass transfer depending on the initial 
mass ratio $q_{\rm i}$ and initial separation. Full line: thermal 
equilibrium radius $R_{\rm 2,e}$ of a star with hydrogen shell burning 
and a non-degenerate He core of mass $M_{\rm He}$ as a function of total 
mass $M_2$ (generalized main sequence). $R_{\rm HL}$ is the maximum 
radius attained at the Hayashi line, $R_{\rm SC}$ the radius at the
Sch\"onberg--Chandrasekhar limit. Dashed lines: Roche lobe radius
$R_{\rm L}$ as a function of $M_2$ for three different values of the 
initial mass ratio $q_{\rm i}$ and the maximum possible orbital separation. 
The dashed lines can be shifted downwards by an arbitrary amount 
subject to the condition that $R_{\rm 2i} \ge R_{\rm SC}$. Dash-dotted line:
mass transfer starting when the secondary has just reached the 
Sch\"onberg--Chandrasekhar limit. This example shows that $q_{\rm crit}$ 
as defined in section 3 depends on the initial orbital 
separation of the binary.}  
\end{figure}

The thermal equilibrium radius $R_{\rm 2,e}$ depends on the relative
mass $M_{\rm He}/M_2$ of the donor's helium core. Figs. 1 -- 3 show 
the so-called `generalized main sequences' of GKW, 
the first schematically, and the latter two for $M_{\rm 2i}
= 3\msun, 5\msun$. For a core-envelope structure to 
be applicable,
the star must have at least finished central hydrogen burning.
Since the mass transfer takes place on the thermal time-scale
of the donor there is little nuclear evolution, and we can regard
$M_{\rm He}$ as fixed. The quantity $R_{\rm 2,e}$ shown in Figs. 1 -- 3
therefore gives the thermal equilibrium radius attained by the star
after transferring varying amounts of its hydrogen envelope. The
evolution of the system is now specified by the initial mass ratio
$q_{\rm i}$ and the radius $R_{\rm 2i}$ of the donor at the onset of mass
transfer. This can lie between the  
maximum radius $R_{\rm B}$ reached during central hydrogen burning
and one almost as large as the value $R_{\rm HL}$ at the Hayashi
line (mass transfer is adiabatically 
unstable if the donor develops a deep convective envelope).
The allowed initial radius range is about a factor 2 for a donor
with $M_{\rm 2i} = 2.5\msun$, increasing to a factor $\sim 6$ for $M_{\rm 2i}
= 5\msun$ (Bressan et al. 1993). 

\begin{table*}
\begin{minipage}{\hsize}%{200mm}
\caption{Outcomes of Case B  evolution with a neutron-star primary} 

\begin{tabular}{llll}\hline
subcase     & intermediate stages  &  $P_{\rm f}$~(d) & final WD
companion \\ \hline
            &                      &          &  \\
low mass    & Hayashi-line LMXB   &  $\sim 10 -
            1000$ & He, obeys mass-period relation\\
            &                      &          \\

early massive, $q_{\rm i} < q_{\rm crit}$ &  Hertzsprung-gap XRB, 
Case C mass transfer  
& $\ga 1000$ & CO, obeys mass-period relation \\
            &                      &       &   \\
early massive, $q_{\rm i} \simeq q_{\rm crit}$
& Cygnus X-2       & $\sim 10$ & CO, overmassive \\
 
            &                      &         &  \\
early massive, $q_{\rm i} > q_{\rm crit}$
& NS + He star, Case BB mass transfer? & $\la 1 - 10$, 
& CO, overmassive \\ \hline

\end{tabular}
\end{minipage}
\end{table*}

We see from the schematic
Fig. 1 that three qualitatively different outcomes of early massive
Case B mass transfer are possible:

1. `small' $q_{\rm i}$, i.e. $M_{\rm 2i}$ only slightly larger than $M_1$. Here
the Roche lobe soon begins to expand, so the curves of $\log R_{\rm L}$ and
$\log R_{\rm 2,e}$ cross before much mass is transferred. 
Mass transfer ends with ignition of central helium burning as this
makes the star shrink somewhat. At this point the secondary still has  
a thick hydrogen envelope and lies on the 
Hayashi line, with the binary having a long orbital period. 
During central helium burning the secondary stays on the Hayashi
line but remains detached. Mass transfer resumes only after central 
helium burning when the star approaches the asymptotic giant branch (AGB).
Mass transfer is again driven by nuclear evolution (double
shell-burning). Since core helium-burning has
ceased this is Case C. Here (unusually) the mass
transfer is adiabatically 
stable, despite the secondary's deep convective envelope,
since the mass ratio is already $\lta 1$. The average mass transfer
rate is again of order $10^{-6}\msun$~yr$^{-1}$, but may become up to
an order of magnitude higher during thermal pulses (Pastetter \&
Ritter 1989). The donor may also lose
large amounts of envelope mass in a wind. If the binary again manages to
avoid common-envelope evolution by ejecting most of the transferred
mass, mass transfer will finally end once the secondary's envelope has
been lost, leaving a very wide (period $\sim 10$~yr) binary containing
a neutron star and a CO white dwarf.

2. `critical' $q_{\rm i}$. We define this as the case where 
the $R_{\rm L}, R_{\rm 2,e}$ curves 
cross at the `knee' in the mass--radius curve, i.e.
with $M_2$ only slightly
larger than $M_{\rm He}$, so the mass transfer depletes almost the
entire envelope. The remnant donor retains a thin hydrogen envelope,
and lies between the main sequence and the Hayashi line on the HR
diagram. Thus the envelope mass is low enough to prevent the star
lying on the Hayashi line, but not so low that the remnant is small
and hot, i.e. to the left of the main sequence. The initial separation
must be small enough that mass transfer starts before central helium
burning, but large enough that it starts only after the donor has 
reached the Sch\"onberg--Chandrasekhar limit.
This limit is defined as the point
where the isothermal helium core has reached the maximum mass which is
able to support the overlying layers of the star, i.e. the point at
which core collapse begins. 
The orbital period is shorter than in case 1.,
but longer than in case 3. below. Cygnus X-2 is an example of this
evolution, viewed at the point where the donor has almost attained its
thermal-equilibrium radius, and mass transfer is well below
the maximum thermal-time-scale rate. This evolution ends with nuclear
evolution of the donor to smaller radii as the 
mass of the hydrogen-rich envelope is further reduced 
by shell burning. 
The system detaches, leaving a helium-star remnant which 
subsequently ignites central helium burning and finally
becomes a CO white dwarf.

3. `large' $q_{\rm i}$. The curves cross only when the hydrogen envelope is
effectively exhausted. The remnant is a helium star and the
orbital period is short. If $M_{\rm He} \lta 0.9\msun$ the helium star
evolves directly into a CO white dwarf. If $M_{\rm He} \ga 1\msun$
this star re-expands during helium shell-burning.
This in turn can give rise to a further phase of 
(so-called Case BB) mass transfer, e.g. Delgado \& Thomas, 1981; 
Law \& Ritter, 1983; Habets, 1985, 1986).

Table 2 summarizes the possible Case B evolutions with a neutron-star
primary. The various outcomes all reflect the general tendency of
larger $q_{\rm i}$ (i.e. larger $M_{\rm 2i}$) to produce greater orbital 
contraction, and thus smaller and relatively less massive remnants 
(their mass is a smaller fraction of a larger $M_{2i}$) and short 
orbital periods. Conversely, larger $R_{\rm 2i}$ in the range 
$R_{\rm B} - R_{\rm HL}$, where $R_{\rm B}$ is the maximum radius
reached during
central hydrogen burning, produces exactly the opposite trends. To model 
Cygnus X-2 one thus needs $q_{\rm i}$ close to the critical value. 

As can be seen from the following arguments the range of possible
solutions is strongly constrained by equation
(\ref{rl}): with the assumption
that $M_1$ remains essentially constant during the
evolution, $M_{\rm 2f}$ is fixed by the observed mass ratio $q_{\rm f} = 0.34
\pm 0.04$. On the other hand, $q_{\rm i}$ is also essentially fixed by the
model assumption that the current donor is close to the end of Case B
mass transfer. This in turn means that the donor is now close to
thermal equilibrium, with its luminosity therefore coming mainly from
hydrogen shell burning.  

\begin{figure}%[ht]
  \centerline{\epsfxsize=1.0\hsize\epsffile{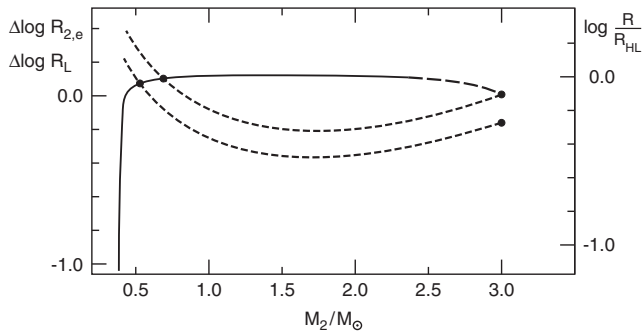}}
\caption{Mass-radius diagram for early massive Case B evolution. Full
(and long dashed) line: generalized main sequence for $3 \msun$ stars 
taken from GKW. Short dashed lines: Roche lobe radius $R_{\rm L}$ as a
function of $M_2$ computed from (3) with $M_1=1.4 \msun,
M_{\rm 2i}=3\msun$ for two values of $R_{\rm 2i}$ ($0.75 R_{\rm HL}$ and $0.55 
R_{\rm HL}$).}
\end{figure}

\begin{figure}%[ht]
  \centerline{\epsfxsize=1.0\hsize\epsffile{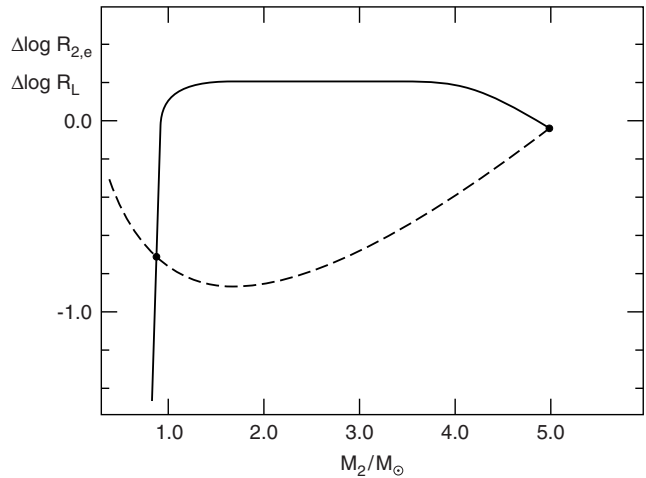}}
\caption{Mass-radius diagram for early massive Case B evolution. 
Full line: generalized main sequence for $5 \msun$ stars taken from 
GKW. Short dashed line: Roche lobe radius $R_{\rm L}$ as a function of
$M_2$ computed from (3) with $M_1 = 1.4 \msun, M_{\rm 2i} = 5
\msun$}
\end{figure}

Use of the generalized main sequence for $3 \msun$ stars given in GKW
(their fig. 7, shown here in Fig. 2) demonstrates that one can 
account for the current state of the donor in Cygnus X-2 if this star
had $M_{\rm 2i} \simeq 3\msun$ 
(i.e. $q_{\rm i} \simeq 2.1$) and $0.55 R_{\rm HL} \la R_{\rm 2i} \la 0.75
R_{\rm HL}$  when mass transfer began, corresponding roughly to 
$0.5 \msun \la M_{\rm 2f} \la 0.7 \msun$.
The numerical results of GKW (their Table 2 and Fig. 4) suggest 
that the current state of the donor is then given approximately by
$M_2 = 0.79\msun, M_{\rm He} = 0.67\msun$ and thus $q_0 = M_{\rm
He}/M_2 = 0.85$; with $T_{\rm eff, 2} = 7060$~K, $L_2 = 126\lsun$ and
$R_2 = 7.5\rsun$. These values correspond to an orbital period
$P \simeq 8.4$~d. From (\ref{orb}) we infer that mass transfer started
at an orbital period $P_{\rm i} \simeq 3.5$~d.
These quantities are close to those given in Table 1, although the
predicted $M_2$ is slightly larger than the observational estimate. 
This may either result from the fact that the chemical composition of 
the models defining the generalized main sequences is characterized by
a step function at $M_{\rm r} = M_{\rm c}$ and thus differs from that of 
evolutionary models, or from the fact that the initial chemical
composition ($X=0.602, Z= 0.044$) and the opacities used by GKW are 
rather outdated. In addition the core mass  
$M_{\rm He} = 0.67\msun$ inferred above
points to an initial mass higher than the
$\sim 3 \msun$ suggested earlier. In fact to fit the observed
value of $0.5 \msun \la M_{\rm 2f} \la 0.7 \msun$, more modern calculations
than those of GKW (e.g. Bressan et al. 1993) yield the required core 
mass if the initial mass was $3.2 \la M_{\rm 2i} \la 4.1 \msun$.

We conclude that the observational data for Cygnus X-2
are well reproduced if we assume it is a remnant of early massive Case
B evolution with $q_{\rm i}$ close to the critical value $\sim 2.3 - 2.9$,
if $M_1 \sim 1.4 \msun$, or $\sim 1.8 - 2.3$, if we adopt the value 
$M_1 \sim 1.8 \msun$ derived by Orosz \& Kuulkers (1998). For the
remainder of this paper we shall adopt $M_1 = 1.4 \msun$.

\section{END PRODUCTS}

The discussion above shows that for both low-mass Case B, and for 
early massive Case B with $q_{\rm i}$ below a critical value ($\sim 2.6$ for 
$M_{2i} \sim 3.6 \msun$), the
evolution leads to a long-period binary with the donor on the Hayashi
line. As is well known, the luminosity and radius of such a star are
fixed by its degenerate core mass $M_{\rm c}$ rather than its total mass $M_2$.
Even though the degenerate core is different in nature in the
two cases, it is possible to give a single formula for the radius, i.e
\be 
r_2 = {3.7\times 10^3m_{\rm c}^4\over 1 + m_{\rm c}^3 + 1.75m_{\rm c}^4},
\protect\label{egg}
\ee
where $r_2 = R_2/\rsun, m_{\rm c} = M_{\rm c}/\msun$ (Joss, Rappaport \& Lewis,
1987). Then using the well-known relation 
\be
P = 0.38{r_2^{3/2}\over m_2^{1/2}}\ {\rm d}
\protect\label{p}
\ee
($m_2 = M_2/\msun$) which follows from Roche geometry, we get a 
relation between $M_{\rm c}, M_2$ and the orbital period $P$ (e.g. King,
1988). 
Once all of the envelope mass has been transferred we are left with 
a wide binary containing a millisecond pulsar (the spun-up
neutron star) in a circular orbit with the white-dwarf core of the 
donor. Since at the end of mass transfer we obviously have 
$m_2 = m_{\rm c}$, such systems should obey the relation
\be
P \simeq 8.5\times 10^4\biggl({m_2^{11/2}
\over [1 + m_2^3 + 1.75m_2^4]^{3/2}}\biggr)\ {\rm d}.
\protect\label{h}
\ee
The timing orbit of the millisecond pulsar allows constraints on
the companion mass, so this relation can be tested by observation.
Lorimer et al. (1995), Rappaport et al. (1995) and Burderi, King \&
Wynn (1996) show that while the relation is
consistent with the data for a majority of the $\sim 25$ relevant
systems, there are several systems (currently 3 or 4; see Fig. 4 
and Table 3) for which the
white dwarf mass is probably too large to fit. While one might
possibly exclude B0655+64 because of its long spin period $P_{\rm s}$ (but
see below), the
other three systems are clearly genuine millisecond pulsars.

\begin{table}

\caption{Binary pulsars with relatively massive WD companions} 

\begin{tabular}{llll}\hline
pulsar &  $P$ (d) & $M_2$ ($\msun$) & $P_{\rm s}$ (ms)\\ \hline
J0621+1002      & 8.32           &   $ > 0.45$ & 28.8 \\
J1022+1001      & 7.81           &   $ > 0.73$ & 16.5 \\
J2145-0750      & 6.84           &   $ > 0.43$ & 16.1 \\
B0655+64        & 1.03           &   $   0.7:$ & 196  \\ \hline 
\end{tabular}

\end{table}

Our considerations here offer a simple explanation for this
discrepancy. If the initial mass ratio $q_{\rm i}$ lies above the critical 
value ($\sim 2.6$ for $M_{2i} \sim 3.6 \msun$), the donor radius will be
less than $R_{\rm HL}$ at the end of early massive Case B mass transfer, 
and the orbital period relatively shorter. When such systems finally 
detach from the Roche lobe, the WD companion is considerably more
massive than expected for the orbital period on the basis of
the Hayashi-line relation (\ref{h}). We see from (\ref{orb}) that
systems with large initial companion masses $M_{\rm 2i}$ ($\ga 4\msun$)
can end as short-period systems. Table 4 and Fig. 4 show the expected
minimum final periods $P_{\rm f}$ and companion masses $M_2$ for various 
$M_{\rm 2i}$, i.e. assuming that mass transfer began with the donor at the 
Sch\"onberg--Chandrasekhar limit. 

\begin{table}

\caption{Short-period end-products of early massive Case B evolution} 

\begin{tabular}{lll}\hline
$M_{\rm 2i}\ (\msun)$     & $P_{\rm f}({\rm min})$ (d)  &  $M_2\
(\msun)$ \\ \hline
4.0                & 6.60      &  0.56 \\
5.0                & 1.25      &  0.79 \\
6.0                & 0.268     &  1.05 \\
7.0                & 0.048     &  1.46 \\ \hline
\end{tabular}

The minimum
final period $P_{\rm f}({\rm min})$ 
is calculated assuming that mass transfer began
at the Sch\"onberg--Chandrasekhar limit. Initial donor masses 
$M_{\rm 2i} \ga 4\msun $ may possibly be ruled
out because of delayed dynamical instability (see text).\\
\end{table}

\begin{figure}%[ht]
  \centerline{\epsfxsize=1.0\hsize\epsffile{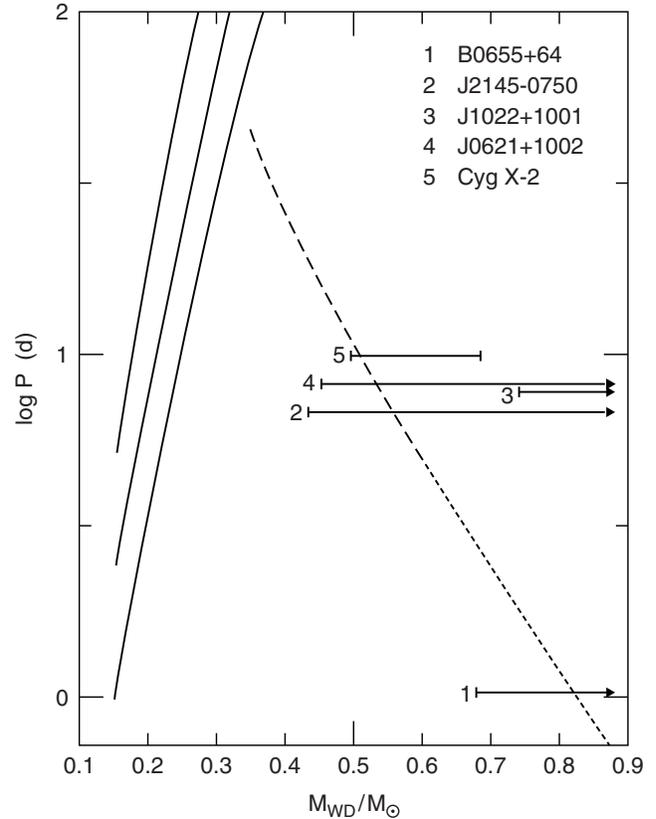}}
\caption{$P - M_2$ diagram for long-orbital-period binary ms-pulsars 
and Cygnus X-2. Full lines: predicted $P - M_2$ relation for 
systems in which the donor star evolved along the Hayashi line (taken
from Rappaport et al. 1995). Dashed line: predicted {\it minimum} orbital 
period for systems in which the white dwarf secondary was formed via
early massive Case B mass transfer. The short-dashed part of this 
curve is realised only if the neutron star can expel the transferred 
mass despite the binary being subject to the delayed dynamical mass
transfer instability. Also shown are the ($P$, $M_2$) values of
the four `discrepant' ms-pulsar binaries listed in Table 3, and the
current position of Cygnus X-2.}
\end{figure}

Table 4 and Fig. 4 show that early massive Case B evolution with $q_{\rm i}$
larger than the critical value can explain the discrepant pulsar
systems of Table 3. Indeed it appears that the process can end with 
very short orbital periods, offering an alternative to the usual 
assumption of a common-envelope phase (Bhattacharya, 1996; Tauris,
1996) The limiting factor for this 
kind of evolution may be the so--called `delayed dynamical instability' 
(Webbink, 1977; Hjellming, 1989). For a sufficiently massive initial 
donor, mass transfer eventually becomes dynamically unstable because 
the adiabatic mass radius exponent of a strongly stripped radiative 
star becomes negative and the donor begins to expand adiabatically in 
response to mass loss. For typical neutron star masses $M_1 \simeq 
1.4\msun$ this would limit $M_{\rm 2i}$ to values $\la (4 - 4.5) \msun$ 
(Hjellming 1989; Kalogera \& Webbink, 1996). Given the
uncertainties in our estimates, this is probably consistent with 
the initial mass $M_{\rm 2i} \ga 5\msun$ needed to explain the current
state of the most extreme discrepant system (B0655+64). The other
three systems can all be fitted with $M_{\rm 2i} \la 4\msun$, so there
is no necessary conflict with the mass limits for the delayed
dynamical instability.

We may now ask about the end states of such an evolution if the
neutron star were unable to eject the mass transferred at the very 
high rates expected once the delayed dynamical instability set in and 
the system went through a common envelope phase instead. Using the
standard prescription for estimating the parameters of a post common
envelope system (e.g. Webbink 1984) with the values for $M_{\rm 2i}$ and 
$M_{\rm 2f} = M_2$ given in Table 4, $M_1 = 1.4 \msun$ and $\alpha_{\rm CE}
\lambda = 0.5$, where $\lambda \sim 0.5$ is a structural parameter and 
$\alpha_{\rm CE}$ the common envelope efficiency parameter  defined
by Webbink (1984), we find that the final orbital period is $0.04 \la 
P_{\rm f}(d) \la 0.2$ if mass transfer set in when the donor was already
near the Hayashi line, and smaller still if mass transfer set in
earlier or if $\alpha_{\rm CE}$ is smaller than unity. On the other hand,
unless $\alpha_{\rm CE}$ is significantly smaller than unity, common
envelope evolution
starting from a system with the donor on the asymptotic giant
branch would end with periods much longer than those of the systems
listed in Table 3 and shown in Fig. 4. 
Table 5 shows the outcome of common--envelope evolution in the two
extreme cases where mass transfer starts at the
Sch\"onberg--Chandrasekhar limit, and on the Hayashi line. Although
common--envelope efficiencies exceeding unity are discussed in the
literature (i.e. energy sources other than the orbit are used to expel
the envelope), the values found for 
$\beta_{\rm CE}=\lambda~\alpha_{\rm CE}$ in
Table 5 show that to produce Cyg X--2--like systems
requires absurdly large efficiencies, even starting from the most
distended donor possible. Thus common envelope
evolution does not offer a promising explanation for these systems.
Their very existence may thus indicate that an accreting neutron star can 
eject mass efficiently even at the very high mass transfer rates
encountered in the delayed dynamical instability. We conclude 
therefore that even very rapid mass transfer 
on to a neutron star does not necessarily result in a common
envelope (cf King \& Begelman, 1999).

\begin{table*}
\begin{minipage}{\hsize}%{200mm}

\caption{Outcomes of common--envelope evolution for early massive Case
B} 

\begin{tabular}{lccccccc}\hline
%\begin{tabular}{llllllll}\hline
Case & $M_{\rm 2i}$ ($\msun$) & $M_{\rm 2f}$ ($\msun$) &  
$M_1$ ($\msun$) & $R_{\rm 2i}$ ($\rsun$)
& $P_{\rm f}({\rm d})$ for $\beta_{\rm CE} = 0.5$ & $\beta_{\rm CE}$
for $P_{\rm f} = 8$~d & $\beta_{\rm CE}$ for $P_{\rm f} = 1$~d \\ \hline

$R_{\rm 2i} = R_{\rm SC}$ 
 &4.0 & 0.56 & 1.4 & 8.5  & 0.0035 & 86.9 & 21.7     \\
 &5.0 & 0.79 & 1.4 & 10.9 & 0.0043 & 75.7 & 18.9     \\
 &6.0 & 1.05 & 1.4 & 14.2 & 0.0054 & 64.8 & 16.2     \\
 &7.0 & 1.46 & 1.4 & 16.7 & 0.0071 & 54.2 & 13.6     \\
 &    &      &     &      &        &      &          \\
$R_{\rm 2i} = R_{\rm HL}$ 
 &4.0 & 0.56 & 1.4 & 41.4 & 0.038  & 17.8 & 4.5      \\
 &5.0 & 0.79 & 1.4 & 71.0 & 0.071  & 11.7 & 2.9      \\
 &6.0 & 1.05 & 1.4 & 109. & 0.12   & 8.5  & 2.1      \\
 &7.0 & 1.46 & 1.4 & 153. & 0.20   & 5.9  & 1.5      \\ \hline

\end{tabular}

The initial system in each case
consists of a neutron star (mass $M_1$) a donor at the onset of
massive Case B evolution (mass $M_{2i}$, core mass $M_{2f}$). In the
upper half of the table mass transfer is assumed to start when the
donor reaches the Sch\"onberg--Chandrasekhar limit, corresponding to
the minimum possible orbital separation. In the lower half of the
table mass transfer is assumed to start only when the donor has
reached the Hayashi line, corresponding to the maximum possible
orbital separation. The parameter $\beta_{\rm CE} = 
\lambda~\alpha_{\rm CE}$. 
\end{minipage}
\end{table*}

We note finally that all of the pulsars of Table 3 have
spin periods much longer than their likely equilibrium
periods (i.e. they lie far from the `spinup line', cf Bhattacharya \&
van den Heuvel, 1991), suggesting that they have accreted
very little mass ($<< 0.1\msun$) during their evolution. This agrees
with our proposal that these systems are the direct outcome of
a super--Eddington mass transfer phase in which almost all the
transferred mass is ejected.

\section{SPACE VELOCITY AND POSITION OF CYGNUS X-2 IN THE GALAXY}

The distance to Cygnus X-2 derived from the observations of Type I
X-ray bursts (Smale 1989) is $d = (11.6 \pm 0.3)$~kpc. From the 
galactic coordinates $l=87.33^{\circ}$ and $b = -11.32^{\circ}$ and
the solar galactocentric distance $R_0 = (8.7 \pm 0.6)$~kpc one derives 
a galactocentric distance for Cygnus X-2 of $d_{\rm GC} = (14.2 \pm
0.4)$~kpc and
a distance from the galactic plane of $z = (-2.28 \pm 0.06)$~kpc. Thus
Cygnus X-2 has a very peculiar position indeed, being not only in the
halo but also in the very outskirts of our galaxy. But
not only is its position peculiar, its space velocity with
respect to the galactic centre is also surprising. 
It can be shown that the observed 
heliocentric radial velocity of $\gamma = (-208.6 \pm 0.8)$ km~s$^{-1}$ 
(Casares et al. 1998) is totally incompatible with prograde rotation 
on a circular, even inclined orbit around the galactic centre (Kolb
et al., in preparation). 
The orbit is either highly eccentric and/or retrograde. In either case
Cygnus X-2 must have undergone a major kick in the past, presumably when the
neutron star was formed in a Type II supernova. Since prior to the
supernova explosion the primary was much more massive ($M_{\rm 1i} \ga 10
\msun$) than the secondary ($M_{\rm 2i} \sim 3.6 \msun$), the latter was
still on the main sequence when the supernova exploded. 
Thus the age of Cygnus X-2 (and the time elapsed since the supernova) is well
approximated by the nuclear time-scale of the secondary, which is $\sim 
4~10^8$~yr for a $\sim 3.6 \msun$ star. This means that Cygnus X-2 must
have gone around the galactic center a few times since its birth or
supernova explosion and that, therefore, its birthplace in the
galaxy cannot be inferred from its current position and velocity.

\section{DISCUSSION}

We have shown that the unusual nature of the secondary star in Cygnus
X-2 can be understood if the system is near the end of a phase of
early massive Case B evolution in which almost all of the transferred
material is ejected. The system is unusual in having had an
initial mass ratio $q_{\rm i} = M_2/M_1$ in a narrow critical range near
$q_{\rm i} \simeq 2.6$; smaller
ratios lead to detached systems with the secondary near the
Hayashi line, and larger ratios
produce binary pulsars with fairly short orbital periods and
relatively massive white dwarf companions. During this evolution, much
of the companion's original mass ($\sim 3 \msun$ for Cygnus X-2) is
transferred and consequently lost on the thermal time-scale $\sim 10^6$~yr of
this star. Evidently
the huge mass loss rate and the short duration of this phase
make it difficult to detect any systems in this state; they would
probably resemble Wolf--Rayet stars of the WNe type (i.e. showing
hydrogen). 

Cygnus X-2 is currently near the end of the thermal time scale mass
transfer phase, so that its mass transfer rate is now well below the
thermal time-scale value, and probably given by the accretion rate.
At $\dot M_{\rm acc} \sim 2\times 10^{-8}\msun$ yr$^{-1}$ (Smale, 1998), this 
is nevertheless one of the highest in any LMXB, making it easily detectable.
Only a full calculation of the evolution, with in particular a
detailed model for the secondary, can predict the duration of the current
phase; this is not any easy task, as this star deviates strongly from
thermal equilibrium during  most of the evolution. But it is clear that 
the mass transfer rate will decline as the remaining few tenths
of a solar mass in the hydrogen envelope are transferred. Cygnus
X-2's long orbital period and
large accretion disc mean that even its current mass transfer rate
only slightly exceeds the critical value
required for a persistent rather than a transient LMXB (cf. King, Kolb
\& Burderi 1996), so the system will eventually become transient. Once the
envelope has been transferred, mass transfer will stop, and the system
will become a pulsar binary with about the current orbital period $P =
10$~d, and a white dwarf companion with a mass which is slightly higher
than that of the companion's
present helium core. Clearly since the present core mass is at least
$0.35\msun$ this $P - m_2$ combination will not obey the Hayashi-line
relation (\ref{h}), so Cygnus X-2 will become another `discrepant'
system like those in Table 3. 

The reasoning of the last paragraph shows that Cygnus X-2 will cease to be a
persistent X-ray binary within the current mass transfer time-scale
$t_{\rm M} =
(M_2 - M_{\rm c})/\dot M_{\rm acc} \sim 10^7$~yr. Its past lifetime as a
persistent source before the current epoch, and its future one as a
detectable transient after it, are both likely to be of a similar order,
although full evolutionary calculations are required to check this. The
fact that we nevertheless observe even one system like Cygnus X-2 strongly
suggests that the birthrate of such systems must be relatively high, i.e.
$\sim 10^{-7}$~yr$^{-1}$ in the Galaxy. Since the binary pulsar
end-products of these systems have enormously long lifetimes, this may
suggest that systems like Cygnus X-2 play a very important role in
providing the Galactic population of millisecond pulsars. 

Cygnus X-2 thus fits naturally into a unified description of
long-period LMXBs in which super--Eddington Case B mass transfer is
efficiently ejected by the neutron star. While the ejection process
can already be inferred for the formation history of Hayashi-line
LMXBs resulting from low-mass Case B evolution (Bhattacharya \& van 
den Heuvel, 1991; Kalogera \& Webbink, 1996), Cygnus X-2 supplies 
the most
powerful evidence that this process must occur. The work of Section 2
shows that it is very hard
otherwise to reconcile the rather low current mass ($M_2 \simeq 0.5 -
0.7\msun$) of the secondary
with its large radius ($R_2 \simeq 7\rsun$) and high luminosity ($L_2
\simeq 150\lsun$). From Section 4 we see that the
orbital period $P = 9.84$~d
is far too long for the system to be the product of common-envelope
evolution, leaving no realistic alternative for driving the required
mass ejection.

\section{Acknowledgments} 

ARK thanks the Max--Planck--Institut f\"ur Astrophysik for its
hospitality during 1998 August, and the U.K. Particle Physics and Astronomy 
Research Council for a Senior Fellowship. HR thanks the Leicester
University Astronomy Group for its hospitality during 1998
November/December, and support from its PPARC Short-Term Visitor grant.

\section{Appendix: X--ray Heating in Cygnus X--2}

In Section 2.2 we considered a stripped subgiant model for Cyg X--2,
and asserted that the observed orbital modulation of the optical flux
($\Delta V_{\rm obs} \simeq 0.3$~mag) would require an extremely low
inclination if one appeals to X--ray heating of the companion to raise
its observed effective temperature to 7330~K. Here we justify this
claim.

We consider a simple picture in which the hemisphere of the
(spherical) companion
facing the neutron star has effective temperature 7330~K, while the
other hemisphere has the Hayashi--line effective temperature
4100~K. We consider the effect of relaxing these assumptions
below. Then viewing the heated face at the most favourable phase
the observer sees hot and cool areas $2\pi R_2^2(1/2 + i/\pi), 
2\pi R_2^2(1/2 - i/\pi)$, where $i$ is the inclination in radians,
with the two expressions reversing at the
least favourable phase. Neglecting limb--darkening, the
ratio of maximum to minimum flux is
\be
{F_{\rm max}\over F_{\rm min}} = 
{(1/2 + i/\pi)B_{\rm hot} + (1/2 - i/\pi)B_{\rm cool}\over
(1/2 + i/\pi)B_{\rm cool} + (1/2 - i/\pi)B_{\rm hot}},
\label{a1}
\ee
where $B_{\rm hot}, B_{\rm cool}$ are the optical surface brightnesses
of the hot and cool regions respectively. Approximating these by
Planck functions at $5500\ \AA$, we find 
$B_{\rm hot}/B_{\rm cool} \simeq 15$. Requiring 
$F_{\rm max}/ F_{\rm min} \la 1.3$ ($\Delta V_{\rm obs} \simeq
0.3$~mag) in (\ref{a1}) shows that $i \la 0.0745\pi$, or $i \la
13.4^{\circ}$, as used in Section 2.2

In reality the heated region would be smaller than a
hemisphere, and its temperature higher than 7330~K in order to produce
an average observed temperature of this value. However relaxing these
limits clearly requires even smaller inclinations than the estimate above,
because the contrast in optical surface brightness between the hot and cool
regions would be even larger than the ratio $\sim 15$ we found above.

\end{document}